**Authors: Andreas Kamilaris** [1,2]**, Ian Cole**[1] **and Francesc X. Prenafeta-Boldú** [3]

[1] CYENS Center of Excellence, Nicosia, Cyprus.

[2] Pervasive Systems Group, University of Twente, Enschede, The Netherlands

[3] Institute of Agriculture and Food Research and Technology (IRTA), Barcelona, Spain.

**Abstract:** Blockchain is an emerging digital technology allowing ubiquitous financial transactions among distributed untrusted parties, without the need of intermediaries such as banks. This chapter examines the impact of blockchain technology in agriculture and food supply chain, presents existing ongoing projects and initiatives, and discusses overall implications, challenges and potential, with a critical view over the maturity of these projects. Our findings indicate that blockchain is a promising technology towards a transparent supply chain of food, with many ongoing initiatives in various food products and food-related issues, but many barriers and challenges still exist, which hinder its wider popularity among farmers and systems. These challenges involve technical aspects, education, policies and regulatory frameworks.

## 1. Introduction

A decade has passed since the release of the whitepaper "Bitcoin: A Peer-to-Peer Electronic Cash System" by the pseudonymous author [1]. This work set basis for the development of Bitcoin, the first cryptocurrency that allowed reliable financial transactions without the need of a trusted central authority, such as banks and financial institutions [2]. Bitcoin solved the double-spending problem (i.e. the flaw associated to digital tokens because, as computer files, can easily be duplicated or falsified), with the invention of the blockchain technology. A blockchain is a digital transaction ledger, maintained by a network of multiple computing machines that are not relying on a trusted third party. Individual transaction data files (blocks) are managed through specific software platforms that allow the data to be transmitted, processed, stored, and represented in human readable form. In its original bitcoin configuration, each block contains a header with a time-stamp, transaction data and a link to the previous block. A hash gets generated for every block, based on its contents, and then becomes referred in the heading of the subsequent block (see Figure 1). Hence, any manipulation of a given block would result in a mismatch in the hashes of all successive blocks.

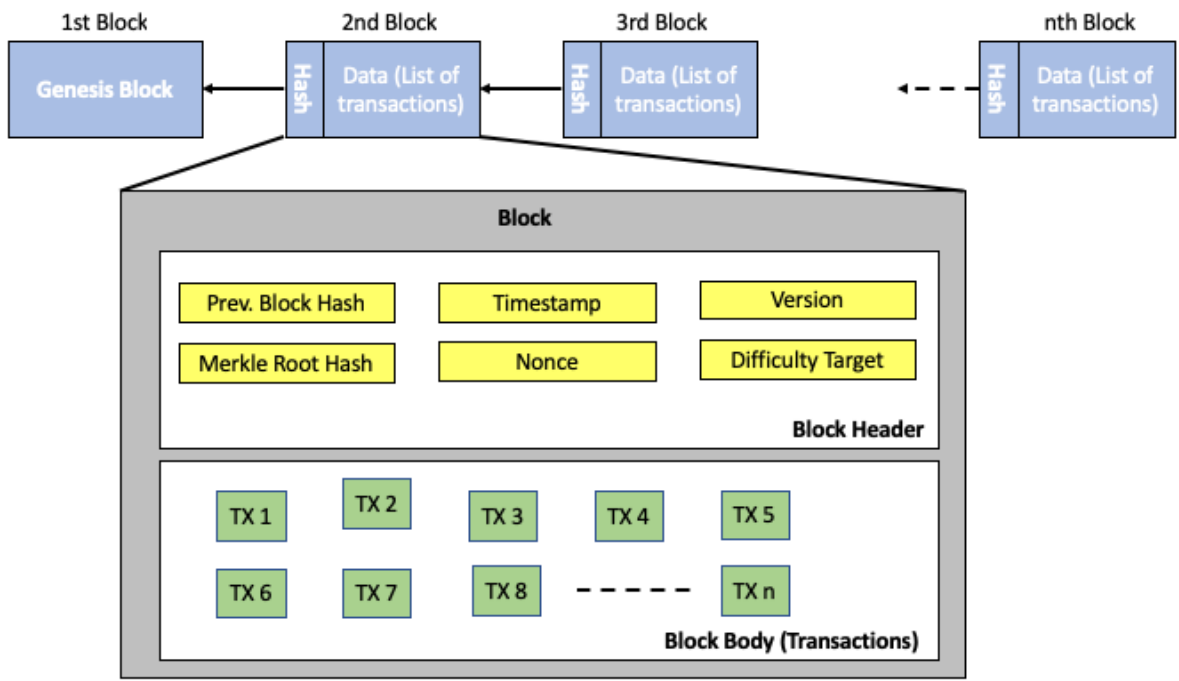

Figure 1: Example of a blockchain containing *n* blocks, in which each successive block contains the hash of the previous block, a timestamp, the transaction information, the nonce number for the mining process and other details needed for the protocol to work.

Every transaction is disseminated through the network of machines running the blockchain protocol, and needs to be validated by all computer nodes. The key feature of a blockchain is its ability to keep a consistent view and agreement among the participants (i.e. *consensus*) [3], even if some of them might not be honest [4]. The problem of consensus has been extensively studied by researchers in the past, however its use in the domain of blockchain has given new stimuli and motivation, leading to novel proposals for design of blockchain systems. The most well-known, used in Bitcoin, is called "Proof of Work" (PoW) and it requires computer nodes, called miners in this case, to solve difficult computational tasks before validating transactions and be able to add them to the blockchain [5]. The first miner to solve the puzzle bundles the block to the chain, which is then validated by the rest, and gets rewarded with newly minted coins plus a small transaction fee.

Common criticism of the PoW include that miners compete continuously in computer power, which leads to increased hardware and energy costs, with the subsequent risks of centralization and high environmental footprint [6], [7]. An alternative consensus approach gaining momentum is called "Proof of Stake" (PoS), and it is about giving the decision-making power to entities who possess coins within the system, putting them "on stake" during transaction approval [5]. In PoS, the nodes are known as the 'validators' and, rather than mining the blockchain, they validate the transactions to earn a transaction fee. There is no mining to be done, as all coins exist from day one. Simply put, nodes are randomly selected to validate blocks, and the probability of this random selection depends on the amount of stake held. Consequently, PoS achieves the same effect of mining (distributed consensus) without the need of expending large amounts of computing power and energy [8]. Other consensus mechanisms include Proof of Elapsed Time (PoET), Simplified Byzantine Fault Tolerance (SBFT), and Proof of Authority (PoA). Hundreds of alternative digital tokens have appeared in the wake of this development, aiming to address some particular weaknesses of the dominant cryptocurrencies, or target a specific domain, such as health, gambling, insurance, agriculture and many others [9]. Blockchain is also being investigated (and in some cases adopted) by the conventional banking system, and nearly 15% of financial institutions are currently using this technology for their transactions [10].

Since 2014 it has increasingly been realized that blockchain can be used for much more than cryptocurrency and financial transactions, so that several new applications are being explored [11]: handling and storing administrative records, digital authentication and signature systems, verifying and tracking ownership of intellectual property rights and patent systems, enabling smart contracts, tracking patient health records, greater transparency in charities, frictionless real-estate transfers, electronic voting, distribution of locally produced goods and, in general, for tracking products as they pass through a supply chain from the manufacturer and distributor, to the final buyer. Such changes are already revolutionizing many aspects of business, government and society in general, but they might also pose new challenges and threads that need to be anticipated. Many of these new applications combine blockchain and distributed ledger technologies (DLTs) with smart contracts and decentralized applications, making third party tampering or censorship virtually impossible [12].

**2. Food Supply Chain**

The food chain worldwide is highly multi-actor based and distributed, with numerous different actors involved, such as farmers, shipping companies, wholesalers and retailers, distributors, and groceries. The main phases characterizing a generic agri-food supply chain are described below [13]:

1. *Production*: The production phase represents all agricultural activities implemented within the farm. The farmer uses raw and organic material (fertilizers, seeds, animal breeds and feeds) to grow crops and livestock. Throughout the year, depending on the cultivations and/or animal production cycle, we can have one or more harvest/yield.
2. *Processing*: This phase concerns the transformation, total or partial, of a primary product into one or more other secondary products. Subsequently a packaging phase is expected, where each package might be uniquely identified through a production batch code containing information such as the production day and the list of raw materials used.
3. *Distribution*: Once packaged and labeled, the product is released for the distribution phase. Depending on the product, delivery time might be set within a certain range and there might be a product storage step (Storage).
4. *Retailing*: At the end of the distribution, the products are delivered to retailers who perform the sale of the product (Retailers). The end-user of the chain will be the customer, who will purchase the product (Customer).
5. *Consumption*: The consumer is the end user of the chain, he/she buys the product and demands traceable information on quality standards, country origin, production methods, etc.

Figure 2 (top section, physical flow) illustrates a simplified version of the food supply system and its main phases and actors. This current system is till date inefficient and unreliable [14]. Exchange of good are based on complex and paper-heavy settlement processes while these processes are not much transparent, with high risks between buyers and sellers during exchange of value. As transactions are vulnerable to fraud, intermediaries get involved, increasing the overall costs of the transfers [15]. It is estimated that the cost of operating supply chains makes up two thirds of the final cost of goods. Thus, there is much space for optimization of the supply chains, by effectively reducing the operating costs. Finally, when people buy products locally, they are not aware of the origins of these goods, or the environmental footprint of production.

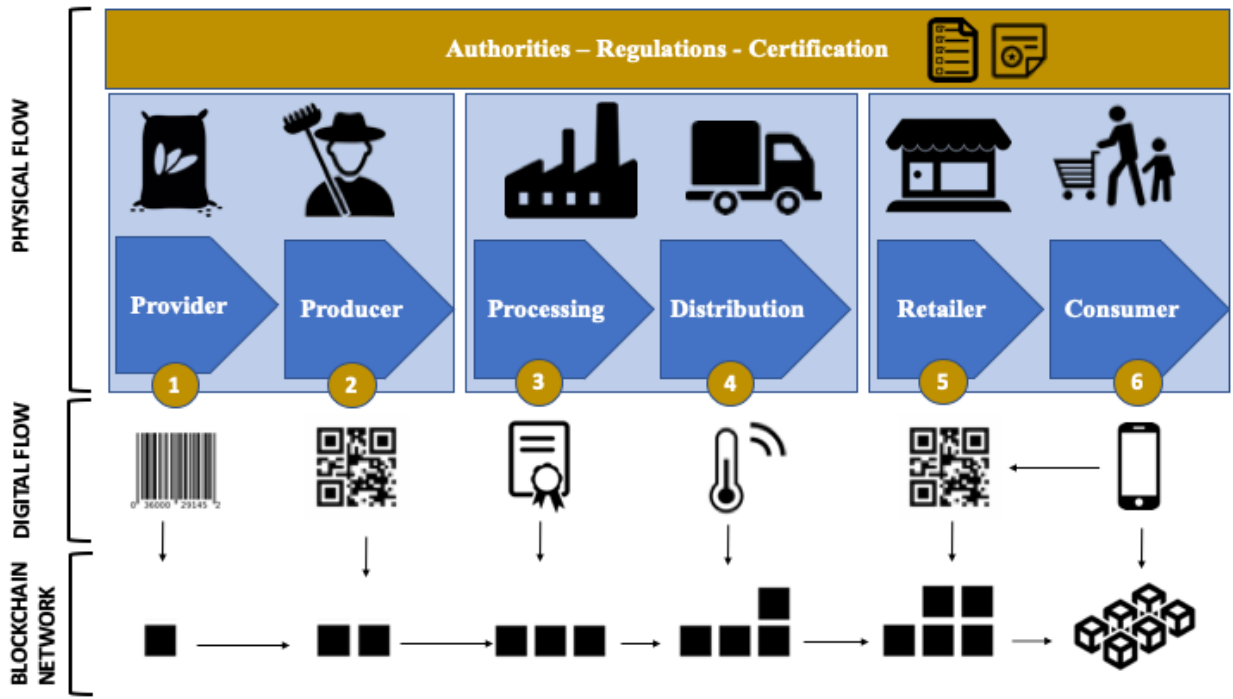

Figure 2: A simplified food supply chain system.

## 3. Blockchain in Agriculture and Food Supply Chain

While the blockchain technology gains success and proves its functionality in many cryptocurrencies, various organizations and other entities aim at harnessing its transparency and fault tolerance in order to solve problems in scenarios where numerous untrusted actors get involved in the distribution of some resource [16], [17]. Two important, highly relevant areas are *agriculture* and *food supply chain* [18], [14].

Agriculture and food supply chains are well interlinked, since the products of agriculture almost always are used as inputs in some multi-actor distributed supply chain, where the consumer is usually the final client [19].

There is evidence that blockchain applications started to become used in the supply chain management soon after the technology appeared [20]. Blockchain in supply chain management is expected to grow at an annual growth rate of 87% and increase from $45 million in 2018 to $3,314.6 million by 2023 [21]. Two survey papers have already been published in scientific journals, aiming to capture this growth and use of blockchain in the sector [22], [23], [24].

As a successful example, in December 2016, the company AgriDigital executed the world's first settlement of the sale of 23.46 tons of grain on a blockchain [25]. Since then, over 1,300 users and more than 1.6 million tons of grain has been transacted over the cloud-based system, involving $360 million in grower payments. The success of AgriDigital served as an inspiration for the potential use of this technology in the agricultural supply chain. AgriDigital is now aiming to build trusted and efficient agricultural supply chains by means of blockchain technology [26]. As another recent example, Louis Dreyfus Co (LDC), one of the world's biggest foodstuffs traders, teamed up with Dutch and French banks for the first agricultural commodity trade (i.e. a cargo of soybeans from the US to China) based on blockchain [27]. According to LDC, by automatically matching data in real time, avoiding duplication and manual checks, document processing was reduced to a fifth of the time.

A simplified example of the digitization of the food supply chain, supported by blockchain technology is depicted in Figure 2. Under the physical flow (top layer), there is the digital flow layer (middle layer), consisting of various digital technologies (i.e. QR codes, RFID, NFC, online certification and digital signatures, sensors and actuators, mobile phones etc.). The Internet/Web serves as the connecting infrastructure. Every action performed along the food chain, empowered by the use of the aforementioned digital technologies, is recorded to the blockchain (bottom layer of Figure 2), which serves as the immutable means to store information that is accepted by all participating parties. The information captured during each transaction is validated by the business partners of the food supply network, forming a consensus between all participants. After each block becomes validated, it is added to the chain of transactions (as Figure 2 shows), becoming a permanent record of the entire process. At every stage of the trajectory of food (defined with numbers 1-6 in Figure 2), different technologies are involved and different information is written to the blockchain, as described below for each of these stages:

1. Provider: Information about the crops, pesticide and fertilizers used, machinery involved etc. The transactions with the producer/farmer are recorded.
2. Producer: Information about the farm and the farming practices employed. Additional info about the crop cultivation process, weather conditions, or animals and their welfare is also possible to be added.
3. Processing: Information about the factory and its equipment, the processing methods used, batch numbers etc. The financial transactions that take place with the producers and also with the distributors are recorded too.
4. Distribution: Shipping details, trajectories followed, storage conditions (e.g. temperature, humidity), time in transit at every transport method etc. All transactions between the distributors and also with the final recipients (i.e. retailers) are written on the blockchain.
5. Retailer: Detailed information about each food item, its current quality and quantity, expiration dates, storage conditions and time spent on the shelf are listed on the chain.
6. Consumer: At the final stage, the consumer can use a mobile phone connected to the Internet/Web or a web application in order to scan a QR code associated with some food item, and see in detail all information associated with the product, from the producer and provider till the retail store.

In this section of the paper, various initiatives have been identified where blockchain technology could be used to solve real-life practical problems at the agricultural supply chain. To identify relevant initiatives, a keyword-based search was performed through the web scientific indexing services *Web of Science* and *Google Scholar*. The following query was used:

*Blockchain AND [Agriculture OR Food OR "Food Supply" OR "Food Supply Chain"].*

Our focus was on *existing* initiatives, projects and case studies, and not on the general potential of blockchain in the field. Based on this search, 59 papers were identified. From these papers, 47 were relevant, in terms of using blockchain technology in food supply chain. To increase bibliography, related work of the initial 59 papers was examined, together with a keyword-based search in popular search engines, increasing the number of relevant identified initiatives to 80. Based on their purpose and overall target/goal, these 80 initiatives were divided into six main categories, as follows:

a) food security (3 projects/initiatives, 4%),
b) food safety (9 projects/initiatives, 11%),
c) food integrity (31 projects/initiatives, 39%),
d) support of small farmers (12 projects/initiatives, 15%),
e) waste reduction, environmental awareness and circular economy (12 projects/initiatives, 15%), and
f) better supervision and management of the supply chain (13 projects/initiatives, 16%).

An analysis of the findings is performed in Section 4. Some of the potential benefits of blockchain are listed in Section 5, while various challenges and barriers for wider adoption are identified and discussed in Section 6.

**3.1 Food Security**

The Food and Agriculture Organization (FAO) defines food security as the situation when "*all people, at all times, have physical, social and economic access to sufficient, safe and nutritious food that meets their dietary needs and food preferences for an active and healthy life*". Achieving this objective has proven to be extremely challenging under humanitarian crises related to environmental disasters, violent political and ethnic conflicts, etc. Blockchain is regarded as an opportunity for the transparent delivery of international aid, for disintermediating the process of delivery, for making records and assets verifiable and accessible and, ultimately, to respond more rapidly and efficiently in the wake of humanitarian emergencies [28]. Examples include digital food coupons having been distributed to Palestinian refugees in the Jordan's Azraq camp [29], via an Ethereum-based blockchain [30], where the coupons could be redeemed via biometric data [31]. At the moment, the project is helping 100,000 refugees.

**3.2 Food Safety**

Food safety is the condition of processing, managing and storing food in hygienic ways, in order to prevent illnesses from occurring to human population. Food safety and quality assurance have become increasingly difficult in times of growing global flows of goods [32]. The Center for Disease Control and Prevention (CDC) claims that contamination because of food causes 48M Americans to become ill and 3,000 to die every year [33], [14]. In 2016, Oceana performed a research on seafood fraud, showing that 20% of seafood is labelled incorrectly [34]. Lee et al. commented that food supply chains are characterized by reduced trust, long shipment distances, high complexity, and large processing times [35]. Blockchain could provide an efficient solution in the urgent need for an improved traceability of food regarding its safety and transparency. As Figure 2 shows, recording information about food products at every stage of the supply chain allows to ensure good hygienic conditions, identifying contaminated products, frauds and risks as early as possible.

Walmart and Kroger are among the first companies to embrace blockchain and include the technology into their supply chains [36], working initially on case studies that focus on Chinese pork and Mexican mangoes [37]. Early results from the studies showed that, when tracking a package of mangoes from the supermarket to the farm where they were grown, it took 6.5 days to identify the origin and the path the fruit followed with traditional methods, whereas with blockchain this information was available in a few seconds [38].

CyberSecurity, a company of Bari, Italy operating in the information technology (IT) security sector, developed the Milk Verification Project prototype [39], in order to tackle food fraud in the dairy supply chain through blockchain technology. This IT tool automates the acquisition and registration of information in the supply chain processes. Another example of application of blockchain in the dairy sector is provided in [40].

The integration of blockchain with Internet of Things (IoT) for real-time monitoring of physical data and tracing based on the hazard analysis and critical control points system

(HACCP) has recently been proposed [41]. This is particularly critical for the maintenance of the cold-chain in the distribution logistics of spoilable food products. As an example, ZetoChain performs environmental monitoring at every link of the cold chain, based on IoT devices [42]. Problems are identified in real-time and the parties involved are notified immediately for fast action taking. Smart contracts are harnessed to increase the safety of sales and deliveries of goods. Mobile apps can be used by consumers to scan *Zeto labels* on products in order to locate the product's history.

The model of Mohan [43] provided an alternative approach to food (chicken) product tracking through blockchain technology, by utilizing existing food quality systems and technology built into the supply chain stages. In this model, all business partners could connect their internal production systems into the blockchain network, due to the standard topology, predefined terminology and sequence of operations, providing assistances and restrictions over traditional tracking systems in terms of food origin and traceability quality, towards enhanced food safety.

Finally, research by George et al. [44] focused on the implementation of a blockchain model in restaurants considering storage time as the major impact variable for various types of fresh pork meat. The prototype system developed captured data from various stakeholders across the food supply chain, segregated it and finally, applied the Food Quality Index (FQI) algorithm to generate a FQI value. The FQI value helped in identifying whether the food was good for consumption based on the specified parameters. FQI value was generated based on the standard storage and handling regulations specified by food safety authorities, and the system checked whether the value derived was within the permissible range.

**3.3 Food Integrity**

Food integrity is about reliable exchange of food in the supply chain. Each actor should deliver complete details about the origin of the goods. Examples of these details have been listed at the beginning of the previous section, and the process is described in Figure 2. This issue is of great concern in China, where the extremely fast growth has created serious transparency problems [41], [45].

Food safety and integrity can be enhanced through higher traceability [46], [32]. By means of blockchain, food companies can mitigate food fraud by quickly identifying and linking outbreaks back to their specific sources [47]. Recent research has predicted that the food traceability market will be worth $14 billion by 2019 [48]. There are numerous examples of companies, start-ups and initiatives aiming to improve food supply chain integrity through the blockchain technology. The most important on-going projects are listed below, based on their scale, their potential impact and the significance of the partners, organizations and/or actors involved.

The agricultural conglomerate Cargill Inc. aims to harness blockchain to let shoppers trace their turkeys from the store to the farm that raised them [49]. Turkeys and animal welfare are considered at a recent pilot involving blockchain [50]. The European grocer Carrefour is using blockchain to verify standards and trace food origins in various categories, covering meat, fish, fruits, vegetables and dairy products [51].

Downstream beer [52] is the first company in the beer sector to use blockchain technology, revealing everything one wants to know about beer, i.e. its ingredients and brewing methods. Every aspect of this craft beer is being recorded and written to the blockchain as a guarantee of transparency and authenticity. Consumers can use their smart phones to scan the QR code on the front of the bottle and they are then taken to a website where they can find relevant information, from raw ingredients to the bottling.

San Domenico roastery [53] adopted blockchain technology to accompany its coffee product with reliable, unmodifiable documentation and guarantee of absolute transparency. Thanks to the blockchain, each step until the sale was recorded and made unmodifiable before being launched to the next step, to ensure unequivocally the quality of the product and the entire production chain associating all the information concerning a coffee product with a univocal QR code. This permitted to access information, news, videos, certifications and images that trace the supply chain up to the final consumer. Figure 3 shows a snapshot of the mobile application after the QR code of an existing coffee product has been scanned, showing information about the origin of the coffee. Once the system was tailored to the company, the cost savings of certification ranged between 70% and 90%, according to

Foodchain [53]. Foodchain is an Italian company that provides traceability services for food supply chains using blockchain technology.

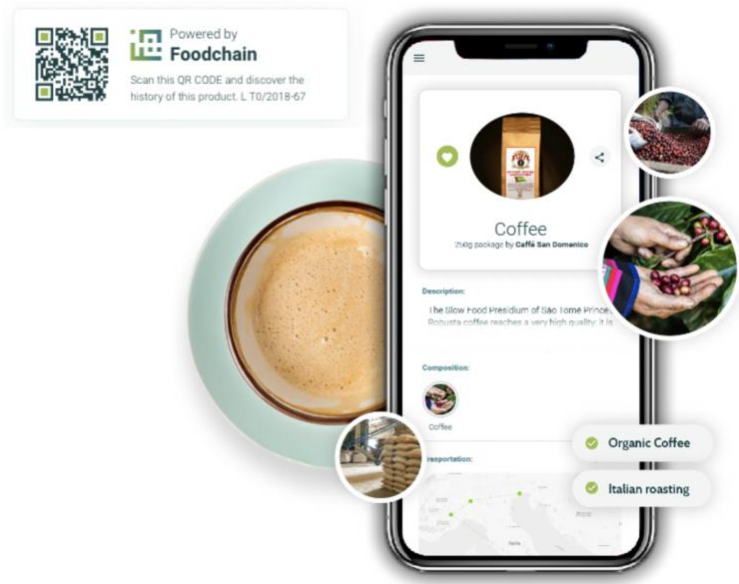

Figure 3: A snapshot of the San Domenico roastery mobile application. After the QR code of an existing coffee product has been scanned, information about the origin of the coffee if shown.

The use of blockchain technology allowed consumers to know the entire supply chain of pasta they were buying in [54]. Framing the QR code printed on the label (similar to the coffee product in Figure 3), the whole supply chain could be identified (i.e. manufacturer, products and flours used, type of drying, transport).

Concerning meat production, "Paddock to plate" is a research project aiming to track beef along the chain of production-consumption, increasing the reputation of Australia for high quality [55]. The project uses BeefLedger as its technology platform [56]. As another example, the e-commerce platform JD.com monitors the beef produced in inner Mongolia, distributed to different provinces of China [57]. By scanning QR codes, one can see details about the animals involved, their nutrition, slaughtering and meat packaging dates, as well as the results of food safety tests. To guarantee to customers that its chickens are actually free-range, the Gogochicken company uses an ankle bracelet to monitor the chickens' movements and behavior via GPS tracking, and this information is then available through the web [58]. The aim of the company is to build trust by documenting the origins of the food. Right now, 100,000 birds have been outfitted with GPS bracelets, but the Shanghai-based company plans to incorporate about 23 million birds into project over the next three years.

The Grass Roots Farmers Cooperative [59] sells a meat subscription box, which uses blockchain technology to inform consumers in a reliable way about the raising conditions of their animals. In the pilot performed, cases of chicken distributed in San Francisco are labeled with QR codes that link to the story of the meat they contain.

Moreover, in April 2017, Intel demonstrated how Hyperledger Sawtooth [60], a platform for creating and managing blockchains, could facilitate traceability at the seafood supply chain. The study used sensory equipment to record information about fish location and storing conditions. Hyperledger is one of the most important initiatives, based on completeness and quality of services and tools, as well as the size of the supporting community and the significance of the members that support the overall project. Hyperledger aims to offer complete solutions towards the business use of the blockchain, and it has been proposed in recent research efforts such as AgriBlockIoT [61]. Hyperledger focuses to the creation of open source frameworks based on the DLT, suitable for enterprise solutions. Two of the most mature Hyperledger frameworks are named *Fabric* (for permissioned blockchain networks) and *Sawtooth* (for both permissioned and permission-less blockchain networks). These two frameworks constitute generic enterprise-grade software, offering support for various smart contract languages and they are used by a wide community of companies, developers and users. In particular, Hyperledger Fabric is backed by IBM. While Hyperledger Fabric is the most well-known and widespread, Sawtooth is the most advanced and heavy-duty, allowing adequate integration with other blockchain frameworks [62]. A demonstrator application based on the Hyperledger Fabric framework was implemented in [63]. The study findings indicated that blockchain technology has entered its maturity phase while on the other hand its adoption in food supply chain operations could add significant value by authenticating critical parameters and providing enhanced traceability.

AgriOpenData Blockchain [46] is an innovative digital technology guaranteeing traceability in the whole agri-food chain for organic and DOCG (Designation of Origin Controlled and Guaranteed) products and in the processing of agricultural products in a transparent, secure, public manner. This integrated system could certificate the quality and the digital identity (provenance, seeding, treatments, crop, IoT, processing, storage, delivery, etc.) of the products assuring authenticity to end-consumers and enhancing the quality of the agri-food business and trust. The transparency of organic food supply chain was also addressed in [64], where instances of smart contracts were created for each physical product, deployed to a blockchain network. Each transaction and event related to a product was validated by peers of the blockchain system, while a token-based mechanism was used to indicate the farmers' reputation with their products. Farmers could place a certification request regarding their products and they could gain reputation tokens for each certification done by peers. An approach that leverages the Ethereum blockchain and smart contracts efficiently to perform business transactions for soybean tracking and traceability across the agricultural supply chain is presented in [65].

In January 2018, the World Wildlife Foundation (WWF) announced the Blockchain Supply Chain Traceability Project [66], to eliminate illegal tuna fishing by means of blockchain. Through the project, fishermen can register their catch on the blockchain through RFID e-tagging and scanning fish. Traceability of tuna is also the focus of Balfegó [67].

Furthermore, ripe.io has created the Blockchain of Food [68], which constitutes a food quality network that maps the food's journey from production to our plate. Ripe.io has recently raised $2.4 million in seed funding in a round led by the venture arm of global container logistics company Maersk [69]. Via the services provided by the OriginTrail company, consumers can see from which orchard the ingredients they cook have grown, the origin and growing conditions of poultry etc. [70]. Also, the project "blockchain for agri-food" developed a proof-of-concept blockchain-based application about table grapes from South Africa [71]. A framework for greenhouse farming with enhanced security, based on blockchain technology, is proposed in [72]. Nestle has recently entered the IBM Food Trust partnership towards food traceability [73], with a pilot based on canned pumpkin and mango. Some research initiatives proposed the combination of blockchain

with other technologies (i.e. IoT, RFID, NFC), in order to increase food traceability. A system based on combining RFID and blockchain technologies is discussed in [74] while a system based on IoT devices and smart contracts is proposed in [75]. Boehm et al. proposed an updated traceability system using blockchain technology combined with Near Field Communication (NFC) and verified users [76].

Canada is currently developing a permissioned blockchain network for the tracking of the cannabis supply chain [77]. By tracking the cannabis chain, Health Canada aims to enforce regulations more easily.

Finally, the blockchain technology is also being assessed to trace the production of non-edible crops that are also very sensitive to integrity issues because of regulation and legal aspects. Figorilli et al. experimented with an implementation of blockchain for the electronic traceability of wood from standing tree to final user, based on RFID sensors and open source technology [78]. The entire forest wood supply chain was simulated in southern Italy (Calabria Region), from standing trees to the final product passing through tree cutting (felling, harvesting, processing) and sawmill process. In this context, the information related to the product quality was integrated with those related to the traceability between RFID architecture and an online information system whose steps (transactions) could be made safe to evidence of alteration through the blockchain.

### 3.4 Small Farmers Support

Small cooperatives of farmers constitute a way to raise competitiveness in developing countries [79]. Via cooperatives, individual farmers are able to win a bigger share of the value of the crops they are cultivating [80]. FarmShare aims to create new forms of ownership of property, cooperation of communities and self-sufficient local economies. It constitutes an evolution of the community-supported agriculture model, taking advantage of the blockchain's potential for distributed consensus, token-based equity shares and automated governance in order to foster greater community engagement while removing some of the managerial burdens [80].

AgriLedger used distributed crypto-ledger to increase trust among small cooperatives in Africa [81]. The authors in [82] proposed a new approach that leads to trusted cooperative applications and services within the agro-food chain, among farmers and other entities of the chain. OlivaCoin is a B2B platform for trade of olive oil, supporting the olive oil market, in order to reduce overall financial costs, increase transparency and gain easier access to global markets [83]. The financial resilience of Kenyan smallholders affected by climate change, through the use of blockchain technology, is discussed in [84], presenting various relevant case studies.

Further, some startups support small farmers by offering tools that increase the traceability of goods, such as Provenance, Arc-Net, Bart.Digital and Bext360. As a recent example, the Soil Association Certification [85] has teamed up with Provenance to pilot technology which tracks the journey of organic food.

The example of Lucena et al. [86] highlighted the advantages obtained with the implementation of a blockchain business network for Brazilian agriculture exports. This platform could help producers track grains stored in warehouses optimizing trading with global exporters allowing for a better flow between the members of the business network and remove the role of some intermediaries in some of the business processes. A similar effort of applying blockchain technology to the cocoa export supply chain of Peru was presented in [87], showing how trust in international buyers is generated via the use of the blockchain. We note here that even medium-size farmers could benefit from blockchain and the aforementioned initiatives, as they form a clearly different category than the large corporations [80]. Cooperatives, on the other hand, might be formed by either small- or medium-size farmers, and can become quite large entities representing tens or hundreds of farmers. Blockchain could be very useful for such cooperatives, because the transparency of information involved could help to solve disputes and conflicts among the farmers in a fairer way for everyone [79], [26]. An example of how blockchain technology could be used for an automatic transaction between a cooperative of farmers (i.e. producers) and a distributor/retailer, via the use of smart contracts, is provided in Figure 4. The figure presents a hypothetical scenario in which a cooperative based in Africa uses a smart contract to facilitate the sale of its cereals' production. The execution of the contract involves the automatic access of the buyer to a storage room, where the crops are stored.

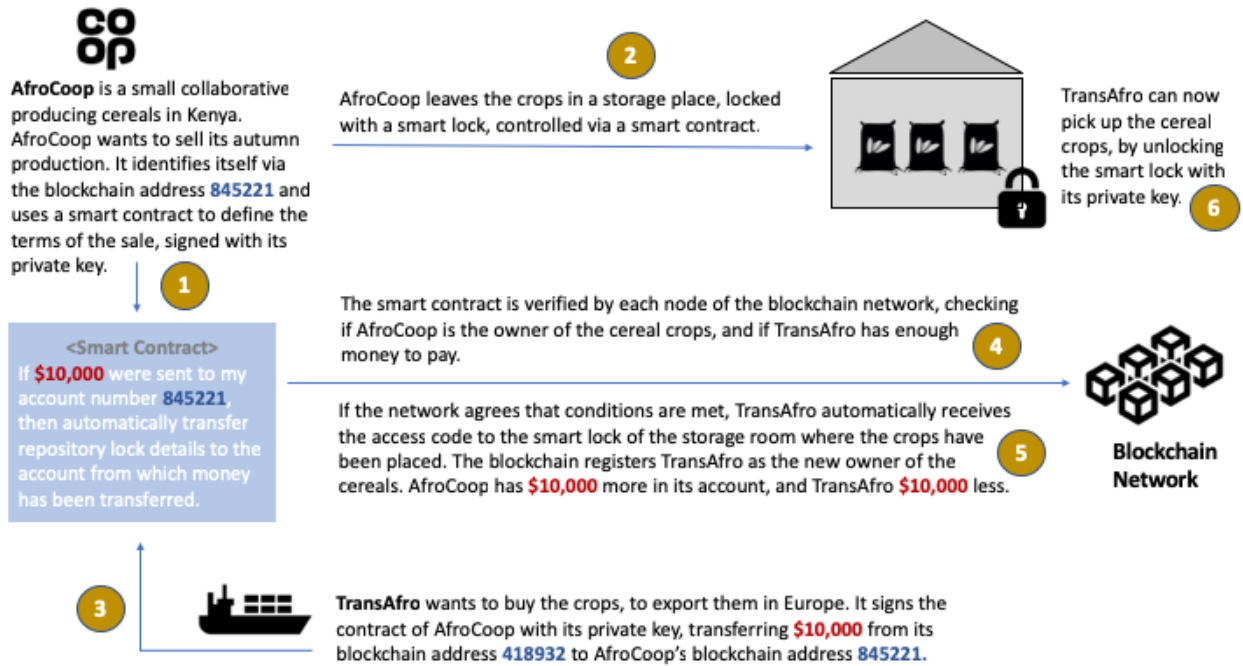

Figure 4: An example demonstrating how a smart contract could be executed in 6 steps, for automating and enhancing trust in transactions involving small farmers and cooperatives of small farmers.

Blockchain could also facilitate insurance programs for securing farmers (i.e. members of the cooperatives) against unpredicted weather conditions that affect their crops or other risks such as natural disasters [88]. The idea behind the ARBOL project is via customized agreements, farmers can receive payments for droughts, floods, or other adverse weather outcomes that negatively affect their crop [89].

Finally, the review by Kim and Laskowski [90] explores blockchain applications across the agricultural sector, beyond the typical finance use cases. A strong focus is given to developing farmers, sustainable agriculture and local economy cooperatives, with pilot programs in Kenya, Myanmar, and Papua New Guinea.

**3.5 Waste reduction, environmental awareness and circular economy**

Various waste management initiatives have incorporated blockchain technology. Worth mentioning is the Plastic Bank [91], a global recycling venture founded in Canada to reduce plastic waste in developing countries – so far Haiti, Peru and Colombia, with plans to extend this year to Indonesia and Philippines. The initiative rewards people who bring plastic rubbish to bank recycling centres, and this reward is provided via blockchain-secured digital tokens. With these tokens, people can purchase things like food or phone-charging units in any store, using the Plastic Bank app [92]. The Plastic Bank initiative seems to be successful till date, with more than one million participants, more than 2,000 collector units and three million kilograms of plastic collected in Haiti since 2014. A company with a mission similar to Plastic Bank is the Agora Tech Lab [93], aiming to promote circular economy initiatives by rewarding responsible behavior.

Another example of the use of blockchain technology is emerging in railway stations. Waste management in French stations has traditionally been chaotic, with hundreds of tones of waste produced each year. A system developed by SNCF subsidiary Arep used the blockchain to allow detailed information to be collected, using Bluetooth to continually update on quantities of each type of waste, which waste managers collected it and how it was being moved around [94]. Blockchain was used to record any actions taken and the overall collection process.

Other commercial solutions using blockchain to improve recycling and sorting of waste produced along the food chain include Recereum [95] and Swachhcoin [96].

An application of blockchain technology in incentivizing the efficient use of rural wastes was proposed by Zhang [97]. The incentive is to trade biomass energy and agricultural products across the waste-to-energy ecosystem. The case study was performed in Changzhi City, Shanxi Province, China considering waste such as crop straw and animal residues.

The authors of [99] proposed a trusted, self-organized, open and ecological food system based on blockchain and IoT technologies that involves all (untrusted) parties of a smart agriculture ecosystem. IoT devices were used to replace manual recording and verification, in order to minimize human intervention. The implementation of smart contract technology was also proposed to fulfil legal requirements.

Moreover, blockchain can help to raise awareness about the environmental characteristics of the food produced. A crucial problem here is the degradation of land, soil and water where food is being produced. In particular, the quality of soil is important towards the realization of the United Nations Sustainable Development Goals (SDG) [100]. In this context, the sustainable development, proper management and rational use of agricultural fields, water resources and soils is of utmost importance [101]. Tracing this information via the supply chain, making it visible to the public, is essential for putting public pressure to producers and policy-makers on the aspect of how the food is produced in a sustainable manner. Finally, a focus on circular economy was taken in some research works. A new model of supply chain via blockchain was proposed in [102], which enables the concept of circular economy and eliminates many of the disadvantages of the current supply chain. A multi-agent system has been designed in order to coordinate all the transactions that take place in the supply chain. The review in [103] presented different case studies on the interactions between blockchain and circular economy in different industrial sectors, including the agri-food systems.

**3.6 Supervision and management**

Blockchain technology can also be harnessed as a credit evaluation system to strengthen the effectiveness of supervision and management in the food supply chain. It can also be used to improve the monitoring of international agreements relevant to agriculture, such as World Trade Organization agreements and the Paris Agreement on Climate Change [14]. The authors in [104] have developed a system, based on the Hyperledger blockchain, which gathers credit evaluation text from traders by smart contracts on the blockchain. Traders' credit can then be used as a reference for regulators, to assess their credibility. By applying blockchain, traders can be held accountable for their actions in the process of transaction and credit evaluation by the regulators. As another example, AgriBlockIoT is a fully decentralized, blockchain-based solution for agri-food supply chain management [13], able to seamless integrate IoT devices producing and consuming digital data along the chain. A similar research effort, combining IoT sensors and cloud technologies was proposed in [82], targeting the management of a grape farm near the City of Skopje, North Macedonia.

Blockchain-based contracts can also mitigate the exploitation of labour in agriculture, protecting workers with temporary agreements and employment relationships in the agricultural sector [105]. When labour agreements become part of the blockchain, it is easier for the authorities to control fairness in payments and also taxation. Coca-Cola has attempted to employ blockchain to sniff out forced labor in the sugarcane sector [106].

Further, the trusted management of water in irrigation communities is an important aspect where blockchain can provide a solution. An implementation of this idea is presented in [107]. A similar implementation, integrating a fuzzy logic algorithm to blockchain for a smart irrigation system is proposed in [108].

Quality measurement and monitoring are also relative aspects, where quality assurance is defined as the avoidance of failures such as delays to final destinations, poor monitoring, and frauds, as well as the assurance that the quality of the products (e.g. crops, meat, dairy) is maintained good along the transfer through the food chain, i.e. good storing conditions, no contamination or impurities etc. Several properties defining a good quality of grains are listed in [109]. The preliminary results in [86] support a potential demand for a blockchain-based certification, which would lead to an added valuation of its selling price around 15% for genetically modified (GM)-free soy in the scope of a business network for grain exports in Brazil. This added valuation would be the outcome of more reliable and efficient quality assurance process on the grains, facilitated by blockchain. Blockchain was also used to record events taking place in the rice value chain, ensuring the security and quality of rice in the transportation process [110]. A conceptual approach for an extension to a mushroom farm distributed process control system with IoT and blockchain integration that allows to collect distributed data on the environmental indicators inherent to mushrooms production is introduced in [111]. The approach moves one step further, complementing the already existent production control system.

Finally, blockchain could be used to manage common resources such as energy, land and water, preventing speculation in the trading of these resources [112]. The work in [113] suggested a system that helps farmers in India to lend agricultural land from landlords easily and securely. The system acted as a bridge between landlords and farmers, using blockchain technology to achieve transparency and security of transactions.

## 4. Analysis of the Findings

Table 1 shows blockchain technology initiatives/projects, in relation to the goods and/or products targeting, based on the examples presented in the previous sections. The last column indicates the objectives for employing blockchain technology at each case. Financial reasons are associated with food traceability in the commercial initiatives. As the table indicates, pilot studies have been implemented in a wide range of different products or at the food supply system as a whole. Some research-oriented studies examined the use of blockchain together with emerging technologies such as IoT, RFID, NFC, QR codes etc., focusing on automation of production and more productivity and transparency [39], [58], [64], [66], [74], [75], [76], [99], [114].

| Goods, Products, Resources | Initiative/Project/Company Involved | Objectives |
|---|---|---|
| Soybeans | LDC [27], Salah et al. [65] | Financial, Faster Operations, Traceability |
| Grains | AgriDigital [26], GEBN study [86] | Financial, Supervision and management |
| Olive oil | OlivaCoin [83] | Financial, Small farmers support |
| Dairy milk | Milk Verification Project prototype [39], Kasten [40] | Food safety |
| Turkeys | Cargill Inc. [49], Hendrix Genetics [50] | Traceability, Animal welfare |
| Mangoes | Walmart, Kroger, IBM [36], [37], Nestle [73] | Traceability |
| Canned pumpkin | Nestle [73] | Traceability |
| Pork | Walmart, Kroger, IBM [36], [37], George et al. [44] | Traceability |

| Sugar cane | Coca-Cola [106] | Supervision and Management |
| --- | --- | --- |
| Beer | Downstream [52] | Traceability |
| Coffee | San Domenico roastery [53] | Traceability |
| Pasta | Aldo Cozzi [54] | Traceability |
| Beef | "Paddock to plate" project [55], BeefLedger [56], JD.com [58] | Traceability |
| Cannabis | Medical Cannabis Tracking (MCT) system [77] | Traceability |
| Chicken | Gogochicken [58], Grass Roots Farmers Cooperative [59], OriginTrail [70], Mohan thesis [43] | Traceability |
| Wood (Chestnut trees) | Figorilli et al. [78] | Traceability |
| Sea-food | Intel [60], WWF [66], Balfegó [67] | Environmental impact, Traceability |
| Table grapes | "Blockchain for agrifood" project [71], Grape farm near the City of Skopje [82] | Experimental feasibility study, Supervision and management |
| Organic food | AgriOpenData Blockchain [46], Basnayake and Rajapakse [64], Soil Association Certification [85] | Financial, Traceability, Small farmers support |
| Cocoa | Chong et al. [87] | Financial, Traceability, Small farmers support |
| Food waste | Plastic Bank (Plastic Bank 2019), Agora Tech Lab [93], SNCF [94], Recereum [95], Swachhcoin [96] | Waste reduction |

| Agricultural by-products, residues and wastes | Crop straw and animal residues [97] | Waste reduction |
|---|---|---|
| Water | Global water assets [112], Management of irrigation communities [107], Smart irrigation system [108] | Supervision and management |
| Rice | Quality of rice in transportation [110] | Supervision and management |
| Mushrooms | Mushroom farm process control system [111] | Supervision and management |
| Agricultural land | Lending of land in India [113] | Supervision and management |
| Food chain in general | AgriLedger [81], FarmShare [80], Carrefour [51], ripe.io [68], OriginTrail [70], AgriBlockIoT [13], Food supply chain prototypes enhanced with other technologies [41], [63]. [75], [76], Ecological food system [99], Local economy cooperatives [90], financial resilience of smallholders affected by climate change [84] | Financial, Traceability, Food safety, Small farmers support, Waste reduction, Supervision and management |
| Supply chain and circular economy | Casado-Vara et al. [102], Kouhizadeh [103] | Waste reduction, environmental impact, circular economy |

Table 1: Goods and products, in relation to projects using blockchain technology and their overall objectives.

## 4.1 Technology

It is interesting to see the underlying technology used by the 80 different projects, initiatives and papers identified through this survey, to empower blockchain-based transactions. The most popular technology adopted was Ethereum (15 projects/initiatives, 19%), followed by Hyperledger Fabric (8 projects/initiatives, 10%). Eight projects preferred to develop their own blockchain solution [26], [42], [51], [57], [68], [70], [83], [107]. From the other initiatives, BigchainDB was employed in [41], the Bitcoin protocol in [49], BeefLedger in [55], Foodchain in [53], the ZhongAn blockchain open platform in [58], Provenance in [59], [85], [114], Hyperledger Sawtooth in [60], the Azure Blockchain Workbench together with Ethereum in [78] and, finally, a combination of Ethereum and Hyperledger Sawtooth [61]. The remaining 38 projects (47%) did not reveal any information about the underlying structure of their blockchain-based solutions. A possible reason could be that some are still in their conceptual stage, as the next section below suggests.

**4.2 Maturity and Sustainability**

Figure 5 depicts the maturity level of the related work as identified through this survey, starting from conceptual stage (17 projects/initiatives, 21%) up to full integration to normal operations of the entity involved (5 projects/initiatives, 6%). As the figure shows, the majority of the projects are either in implementation phase (23 projects/initiatives, 29%) or in a proof-of-concept stage, through small pilot studies (18 projects/initiatives, 22%). Research-based projects tend to reach the level of a small pilot study only, most of them being at a conceptual or implementation stage. All 8 projects/initiatives (10%) that develop large-scale case studies are supported and ran by big companies. With large-scale studies we refer to hundreds of thousands of goods/products involved, interaction with thousands of consumers, and/or involvement of tens to hundreds of intermediate actors in the supply chain. The fact that only 5 initiatives have reached the phase of a complete integration to normal operations [28], [29], [52], [53], [91], indicates that blockchain technology is still being studied by companies and organizations, perceived mostly as an experimental new tool and as an emerging technology with certain potential. It is also likely that companies perform pilot studies involving blockchain for marketing reasons (due to the hype of this technology) or for the possibility of a competitive advantage in the future. Finally, we note that we could not record the maturity level for 11 research papers, since no implementation or deployment details were provided. We strongly believe that they belong to conceptual stage.

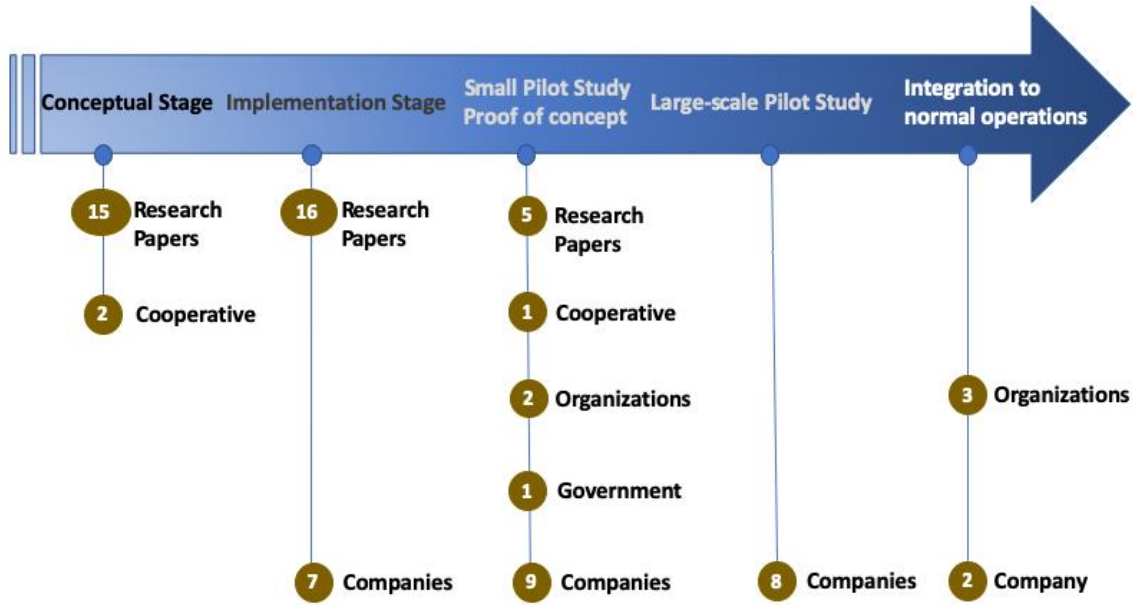

Figure 5: Maturity level and number of projects, initiatives and research papers as identified in this study.

Finally, it is worth investigating whether the aforementioned projects/initiatives are still running, or whether they have stopped and/or failed. This would be a key indicator of the economic viability of blockchain-empowered projects. Unfortunately, it is hard to address this question because most of the initiatives started quite recently, i.e. in 2016 (7 projects/initiatives, 14%), in 2017 (12 projects/initiatives, 24.5%), or in 2018 (22 projects/initiatives, 45%). Due to the small lifetime, most projects are on-going and this makes their assessment difficult. Based on our research, taking into account updates about each project, news articles and any other recorded activity, we suspect that 7 out of the 29 commercial initiatives (24%, including governmental and NGO-based ones, excluding research papers) might have become inactive. These initiatives are the following: [28], [88], [59], [66], [67], [80], [85], [94], [95]. This percentage of possible fallouts is definitely large, and it might be an indication of the overall complexity of the blockchain technology, or the immaturity of the market for complete integration to companies' everyday operations. It could also show that some companies/organizations have finished their pilots, and they are still studying the possibility of massive adoption. Time will show if the latter is the case. Our preliminary findings are in line with the findings in Behnke et al. [114], which mentions that "*use of blockchain remains limited despite its promises*".

**4.3 Potential Benefits**

Blockchain technology offers many benefits, as it can provide a secure, distributed way to perform transactions among different untrusted parties [115], [116], [32]. This is a key element in agriculture and food supply chains, where numerous actors are involved from the raw production to the supermarket shelf [117], [14]. To improve traceability in value chains, a decentralized ledger helps to connect inputs, suppliers, producers, buyers, regulators that are far apart, who are under different programs, different rules (policies) and/or using different applications [35]. Via smart contracts, manufacturers can develop scalable and flexible businesses at a lower cost, and the overall effectiveness of manufacturing services can be improved [118].

Blockchain has the potential to monitor social and environmental responsibility, improve provenance information, facilitate mobile payments, credits and financing, decrease transaction fees, and facilitate real-time management of supply chain transactions in a secure and trustworthy way [35], [97], [111]. In the case of an outbreak of an animal or plant disease, contaminated products could be traced more quickly [14]. Blockchain could even be used to make agricultural robotic swarm operations more secure, autonomous and flexible [119]. It could be used for the easier global distribution and trade of botanical material used as medicines, health foods, in cosmetics and other applications, i.e. to empower high-value botanical value chains [120]. The use of blockchain could also improve the management of nitrogen in crops [98]. In the future, animal and crop farmers could trade organic fertilizers through a blockchain platform.

In particular, blockchain seems very suitable to be used in the developing world, in relation to small farmers' support. Other scenarios could involve finance and insurance of rural farmers [79], financial exploitation of rural waste [97], as well as facilitation of transactions in developing countries. Cash transactions lack traceability, which ultimately hinders the ability of small- and medium-sized enterprises in developing countries, to access credit and new markets and to grow. The blockchain introduces a new method of accounting for value transfers that minimizes uncertainty and disintermediates the exchange of value with a decentralized and shared ledger, functioning as a digital institution of trust, with reduced (if any) transaction costs [14]. Although small farmers produce more than 80% of goods in developing countries, in most cases they do not have support of services such as finance and insurance [79]. Blockchain could also be used to fight corruption and the insufficient environmental, social and economic regulatory frameworks in these countries [121]. More examples on how blockchain could help empowering the poor in developing countries are listed in [122], with focus on tracking climate finance, results tracking, climate adaptation, financial inclusion, and identity.

Concerning the developed world, existing problems such as unfair pricing and the influence of big companies have historically limited the environmental/economic sustainability of smaller farms. Blockchain could help in a fairer pricing through the whole value chain. An example of how blockchain could be used for record keeping of water quality data along a catchment area is discussed in [123].

Moreover, the potential transparency provided by blockchains could facilitate the development of trading systems that are based on reputation. Reputation, as we have witnessed from various other trading systems where it has been used (e.g. eBay, Alibaba), improves the behavior of participating parties and increases their reliability, responsibility and commitment [124], [17].

Further, there is the potential benefit of increasing consumer awareness and empowerment, considering that the consumer is the market driving force. Consumer increased awareness would put pressure for more transparent, sustainable, safe and fair practices in food production. Since consumers are overwhelmed by the amount and complexity of certification labels, blockchain technology seems to have positive influences on consumers' purchasing decisions [125]. Finally, the case study performed in [126] shows that the cost of implementing a blockchain is highly sustainable when compared with the resulting benefits.

## 5. Challenges and Open Issues

There are various challenges for the wider adoption of blockchain technology, which are mentioned in related work under study and also in relevant survey and position papers *[21]*, *[46]*, *[127]*, *[20]*, *[128]*, *[116]*, *[129]*. Table 2 lists potential benefits and existing barriers for the use of blockchain in agriculture and the food supply chain, as identified in the previous sections, as well as in *[21]*, *[116]*. These are still aspects that should be deepened in the food sector to generate a more robust blockchain architecture and ameliorate the themes already treated in this review *[23]*. A case study in the Netherlands revealed that SME lack the required size, scale or know-how needed, in order to invest in blockchain by themselves *[71]*. Eighteen boundary conditions categorized in business, regulation, quality and traceability categories have been identified in *[114]*. Boundary conditions should be met before blockchain can be used. Some boundary conditions were found in all supply chains, whereas others were found to be supply chain specific. Blockchain technology requires standardization and data governance. Blockchain use requires organizational transformations.

| **Opportunities and potential benefits** | **Challenges and barriers** |
| --- | --- |
| Traceability in value chains | SME have difficulties in adopting the technology |
| Support for small farmers | Information infrastructure might prevent access to markets for new users |
| Finance and insurance of rural farmers | Lack of expertise by small SME |
| Facilitation of financial transactions in developing countries | High uncertainties and market volatility |
| Fairer pricing through the whole value chain | Limited education and training platforms |
| A useful platform in emission reduction efforts | No regulations in place |
| Consumer awareness and empowerment | Lack of understanding among policy makers and technical experts |
| More informed consumer purchasing decisions | Open technical questions and scalability issues (e.g. latency of transactions) |
| Increased sustainability and reduction of waste | Digital divide among developed and developing world |
| Reduced transaction fees and less dependence on intermediaries | Decline of cryptocurrencies in market share and high volatility (reputation issues) |
| More transparent transactions and less frauds | Cost of computing/IoT equipment required |
| Better quality of products, lower probability for foodborne diseases | Design decisions might reduce overall flexibility |
| | Privacy issues |
| | Some quality parameters of food products cannot be monitored by objective analytical methods, especially environmental indicators |
| | Ownership of infrastructure and maintenance |
| | Distribution of profits and advantages |
| | Certification of the inserted data |

Table 2: Potential benefits and existing barriers for the use of blockchain in agriculture.

## 5.1 Accessibility

Blockchain needs to become more accessible and this is a big challenge considering that the underlying digital technology can become increasingly complex, as more components are integrated into blockchain (IoT, RFID, sensors and actuators, robots, biometric data, big data, 5G, edge computing etc.) [74], [78], [75], [99], [107], [118], [130], [131], [132]. In fact, in order to be functional, blockchains must rely on external systems to obtain accurate information from the real-world. These are the so-called *oracles* that connect the physical and digital worlds, and usually come from automated sensor readings (i.e. hardware oracles), datasets from the web applications (i.e. software oracles), and manual records (i.e. human oracles). However, the necessity of such third-party intermediaries might compromise the blockchain building of decentralized trust. Substantial research is being carried out on how to tackle the *oracle problem* in blockchain, particularly for finance and smart contract-related applications. The proposed solutions generally rely on developing decentralized and consensus-based oracle solutions, and novel methods of authenticating oracle data.

While blockchains can connect complex global supply chains, the information infrastructure required to operate and maintain the system might prevent access to markets for new users or food suppliers. The systems could, in effect, become a technical barrier to trade, thus reducing market competition and access [116].

Moreover, there is a general lack of awareness and skills on blockchain technology [128], while training platforms are still limited [25]. Besides policy-makers, capacitation on the blockchain technology is also fundamental for the food value chain stakeholders. Conceptual metaphors for understanding and accepting blockchain are discussed in [132]. Various startups have been working in developing software to make blockchain technology easier for farmers to use, such as 1000 EcoFarms [133], which has aggregated all the important blockchain processes relevant to food, farming and agriculture, using FoodCoin as the proposed ecosystem [134]. OriginChain is a software system that restructures the current central database systems with blockchain [135].

**5.2 Governance and Sustainability**

Despite the rather long list of initiatives presented in this review, convincing business cases are still scarce, due to large number of uncertainties involved and the early stages of the technology. This observation was made also in a relevant survey [20]. Hence, the long-term impact of blockchain on governance, economic sustainability, and on social aspects still needs to be assessed. Some authors have pointed out that an excess of information transparency and the immutability of the data stored in blockchains might bring new challenges for the performance of supply chains [127]. On the one hand, permanent data visibility might compromise privacy issues and could eventually strengthen the surveillance power of centralized entities. On the other hand, large corporations might implement private and permissioned blockchains that could underpin oligopolistic practices [116]. The definition of the economic models which could be applied in order to self-feed the supply-chain and the relative blockchain infrastructure need to be carefully defined [23].

Paradoxically, blockchain has also been described as a potentially *deskilling* technology for workers and organizations [127]. The increased automation of tasks and procedures throughout supply chains and the elimination of transaction intermediaries might reduce significantly the human intervention, with the consequent loss of skilled jobs. The margin for human intervention in blockchain-managed supply chains could be reduced significantly. However, we must consider that such phenomena have occurred in all previous technological revolutions, which have in turn demanded new skills and capacities at the labor market.

The distribution of the advantages generated by the cryptocurrency (e.g. profits) is an important aspect that needs to be carefully considered, especially when untrusted partieis are involved *[23]*. The strategy/technology that could be applied in order to take advantage of the blockchain in the certification of the inserted data is also important *[23]*.

Finally, it is worth adding that the quality parameters of food products (being more transparent to the consumer by means of the blockchain) justify in many cases higher prices. Therefore, they are often in the focus of food fraudsters [32], thus governance is important also in this aspect.

**5.3 Regulation**

Policy development and regulation in relation to blockchain practices is both a necessity and an important barrier for its wider adoption [128], [116], [129]. As cryptocurrencies form the most complete to date global blockchain study case [136], the current experience of analyzing these cryptocurrencies indicates that they are vulnerable to speculators and their price has large fluctuations almost daily. The recent decline in market share and high volatility of the financial value of the most popular cryptocurrencies reduces the overall trust of the public in the underlying blockchain technology of cryptocurrencies, thus having a negative psychological effect on its reputation [137]. Hence, without some form of regulation, cryptocurrencies are not trustful to be used yet in food supply chains as a complete solution. The absence of regulation makes this problem persistent.

A lack of (common) understanding among policy makers and technical experts still exists on how blockchain technology and transactions based on some currency should be used

*[25]*. The primary realization of the infrastructure with the different smart contracts and the responsible entities (governmental or certified third part), respecting regulations in place and action constitutes an important challenge *[23]*.

## 5.4 Technical Challenges and Design Decisions

There are many design decisions that affect the existing blockchains or the ones under development (e.g. [26], [81], [80], [68], [70]). For example, shall they be permissioned (i.e. participants are trusted), permission-less, open (i.e. everyone can join) or closed systems [138]? Who should own the blockchain [116]? Observing the existing permission-less blockchains, the latency of transactions might be several minutes up to some hours to finish, until all participants update their ledgers and the smart contracts become publicly accessible. Such design decisions affect the operation of the blockchain system and this creates some lack of flexibility which, under certain circumstances, might make blockchain solutions less efficient than the equivalent conventional centralized approaches. The ownership of the maintenance duty of the infrastructure is another important consideration [23].

Moreover, some of the quality parameters of food products can be monitored by objective analytical methods, but not all of them [32]. Some parameters, especially environmental ones [101], are difficult to include, assess and audit. A preliminary system approach to adopt blockchain in supply chain management for better quality was introduced in [139]. The system was constructed on four layers settled both the technological and supply chain complexity in adopting blockchain in combination with IoT devices for the supply chain management. Improvements on the quality assurance process in a standard Grain Exporters Business Network (GEBN) were reported also in [86]. The product quality of a forest wood supply chain was made available on a blockchain-based online information system in [78]. AgriOpenData Blockchain [46] promises the enhancement of the quality of the agri-food business and trust.

Further, existing blockchain protocols face serious scalability obstacles [140], [116], [129] since the current processing of transactions is limited by parameters such as the size and interval of the transaction block [20]. Mao et al. tried to address this problem by helping users find suitable transactions and improve transaction efficiency [141]. The majority of the proposed blockchain-based frameworks were only tested on a limited scale in a "laboratory" environment. Although blockchain offers advanced security, there are high risks related to loss of funds, just because the account owner might have lost accidentally the private keys needed to access and manage the account.

Privacy issues are also important [128], [129]. Since every transaction is recorded on a common ledger, users can be identified by their public keys. Although this aspect ensures transparency and helps to build trust, at the same time it does not protect users' privacy. This privacy is particularly important in the food supply ecosystem, since many actors are competitors with each other. Thus, maintaining a certain level of privacy is an existing challenge of blockchain technologies. Various methods for privacy protection in blockchain systems are discussed in the survey of [142].

Finally, various aspects of different data standards among different stages in the supply chain when using a decentralized blockchain network were addressed in [143], where an ontology-based blockchain modelling approach was introduced, with the integration of IoT devices for data capturing and data sharing for supply chain provenance. This blockchain technology was built upon Internet technologies, using a Web browser as a natural interface. This could be an early sign that blockchain-driven initiatives in food supply systems would embrace the IoT and the relevant concept of "Web of Things" [144], [145].

**5.5 Digital Gap Between Developed and Developing Countries**

As mentioned in the previous subsection, the farmers need to effectively understand blockchain before adopting it [20]. However, the priority for farmers in many parts of the world is subsistence, so that they need to dedicate their efforts in farming and have no expertise in cutting edge technologies. Since blockchain technologies require a high degree of computing equipment (i.e. in some blockchain systems, such as permissionless ones) [128], it is difficult to find these resources in developing countries. Hence, there seems to be a gap among the developed and developing world, in respect to digital competence and access to the blockchain technology [146]. Many of the bibliographic sources come from developed countries with a well-organized and wealthy primary sector (i.e. the USA,

Australia, Europe, etc.). This digital divide was also observed in the use of big data in agriculture [147]. Some authors do make the important observation that most of the current projects are in developed countries, but no significant questions are raised around this in their conclusion. Since blockchain is being constantly referred to as solving many developing world challenges, asking '*why the gap?*' is an important question, and a legitimate area for future research. Figure 6 illustrates the number of blockchain experiences at the public sector in various countries around the world [148].

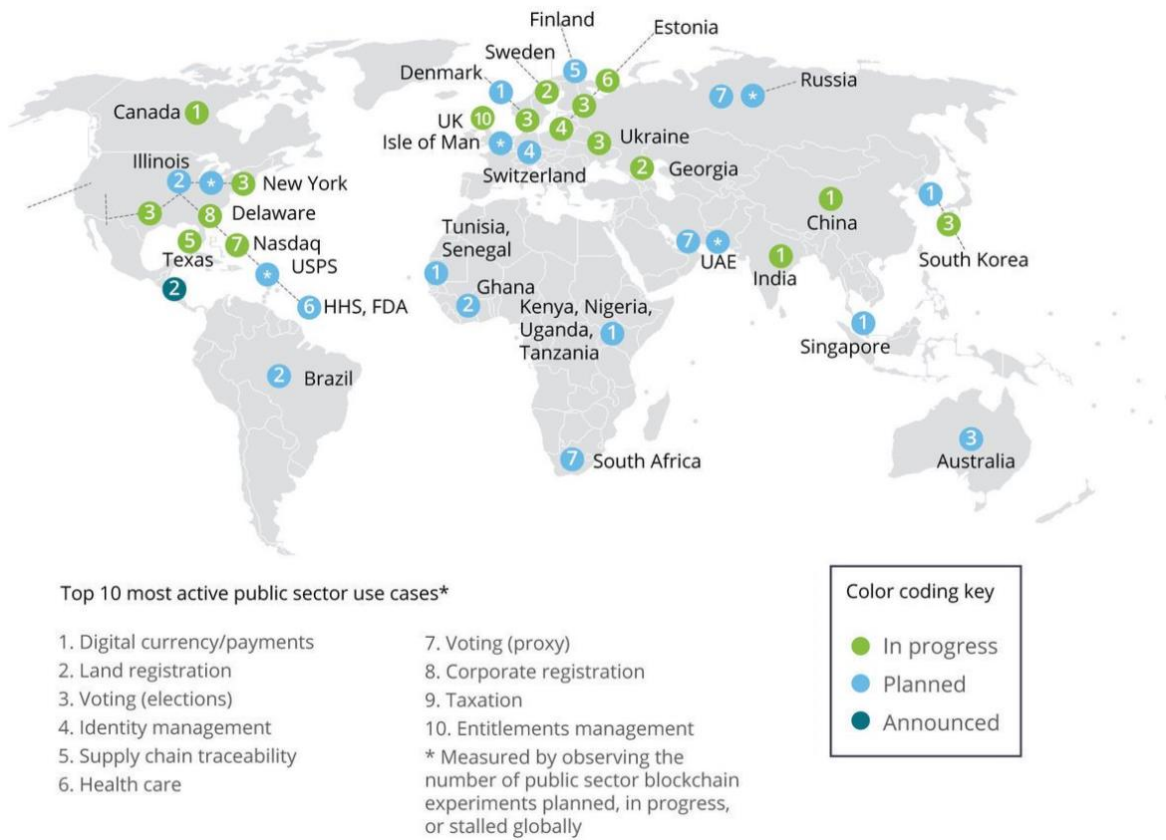

Figure 6: Blockchain in the public sector in 2017 (Source: *[148]*, appropriate permissions have been obtained from the copyright holders of this work).

It seems indeed that most of the on-going experiments happen in developed regions. Considering that blockchain might be an important opportunity for small farmers, developmental aid should focus on training and technology transfer to the farmers in developing areas with the view of bringing actual solutions to the specific conditions that restrain their socioeconomic progression.

## 6. Conclusion

This book chapter demonstrates that blockchain technology is already being used by many projects and initiatives, aiming to establish a proven and trusted environment to build a transparent and more sustainable food production and distribution, integrating key stakeholders into the supply chain. Yet, there are still many issues and challenges that need to be solved, beyond those at technical level.

To reduce barriers of use, governments must *lead by example* and foster the digitalization of the public administration. They should also invest more in research and innovation, as well as in education and training, in order to produce and demonstrate evidence for the potential benefits of this technology. Gupta [149] discussed the possible transition of governments towards the use of the blockchain, noting the fact that governments and their relevant departments should observe and understand the particular "pain points", addressing them accordingly.

From a policy perspective, various actions can be taken, such as encouraging the growth of blockchain-minded ecosystems in agri-food chains, supporting the technology as part of the general goals of optimizing the competitiveness and ensuring the sustainability of the agri-food supply chain, as well as designing a clear regulatory framework for blockchain implementations.

The economic sustainability of the existing initiatives, as they have been presented in this chapter, still needs to be assessed and the outcomes of these economic studies are expected to influence the popularity of the blockchain technology in the near future, applied in the food supply chain domain.

Summing up, blockchain is a promising technology towards a transparent supply chain of food, but many barriers and challenges still exist, which hinder its wider popularity among farmers and food supply systems. The near future will show if and how these challenges could be addressed by governmental and private efforts, in order to establish blockchain technology as a secure, reliable and transparent way to ensure food safety and integrity. It is very interest to see how blockchain will be combined with other emerging technologies (big data, robotics, IoT, RFID, NFC, hyperspectral imaging, 5G, edge computing etc.), towards higher automation of the food supply processes, enhanced with full transparency and traceability.

**Acknowledgments**

Andreas Kamilaris has received funding from the European Union's Horizon 2020 research and innovation programme under grant agreement No 739578 complemented by the Government of the Republic of Cyprus through the Directorate General for European Programmes, Coordination and Development.

**Chapter References**

[1] S. Nakamoto, "Bitcoin: A peer-to- peer electronic cash system," 2008. [Online]. Available: https://bitcoin.org/bitcoin.pdf.

[2] F. Tschorsch and B. Scheuermann, "Bitcoin and beyond: A technical survey on decentralized digital currencies," *IEEE Communications Surveys & Tutorials,* vol. 18, no. 3, pp. 2084-2123, 2016.

[3] S. Bano, "Consensus in the Age of Blockchains," arXiv preprint arXiv:1711.03936, 2017.

[4] M. Castro and B. Liskov, "Practical Byzantine Fault Tolerance," *OSDI,* vol. 99, pp. 173-186, 1999.

[5] I. Bentov, A. Gabizon and A. Mizrahi, "Cryptocurrencies without proof of work," in *International Conference on Financial Cryptography and Data Security*, Berlin, Heidelberg, 2016.

[6] J. Becker, D. Breuker, T. Heide, J. Holler, H. P. Rauer and R. Böhme, "Can We Afford Integrity by Proof-of-Work? Scenarios Inspired by the Bitcoin Currency," in *Economics of Information Security and Privacy*, Berlin, Heidelberg, 2013.

[7] M. Krause and T. Tolaymat, "Quantification of energy and carbon costs for mining cryptocurrencies," *Nature Sustainability,* vol. 1, pp. 711-718, 2018.

[8] BitFury Group, "Proof of Stake versus Proof of Work," White Paper, 2015.

[9] Coinmarketcap, 2017. [Online]. Available: http://www.coinmarketcap.com.

[10] IBM, "Leading the Pack in Blockchain Banking: Trailblazers Set the Pace," IBM Institute for Business Value, The Economist Intelligence Unit, 2017.

[11] S. Tayeb and F. Lago, "Blockchain technology: between high hopes and challenging implications," MENA Business Law Review First quarter, 2018.

[12] V. Buterin, "The Problem of Censorship," 2015. [Online]. Available: https://blog.ethereum.org/2015/06/06/the-problem-of-censorship/ .

[13] M. Caro, M. Ali, M. Vecchio and R. Giaffreda, "Blockchain-based traceability in Agri-Food supply chain management: A practical implementation," in *IoT Vertical and Topical Summit on Agriculture-Tuscany (IOT Tuscany)*, 2018.

[14] M. Tripoli and J. Schmidhuber, "Emerging Opportunities for the Application of Blockchain in the Agri-food Industry," *FAO and ICTSD: Rome and Geneva,* Vols. Licence: CC BY-NC-SA 3 , 2018.

[15] M. Lierow, C. Herzog and P. Oest, Blockchain: The Backbone of Digital Supply Chains, Oliver Wyman, 2017.

[16] S. Manski, "Building the blockchain world: Technological commonwealth or just more of the same?," *Strategic Change,* vol. 26, no. 5, pp. 511-522, 2017.

[17] S. Sharma, "Climate Change and Blockchain," ISO 690, 2017.

[18] D. Dujak and D. Sajter, "Blockchain Applications in Supply Chain," *SMART Supply Network,* pp. 21-46, 2019.

[19] A. Maslova, "Growing the Garden: How to Use Blockchain in Agriculture," 2017. [Online]. Available: https://cointelegraph.com/news/growing-the-garden-how-to-use-blockchain-in-agriculture.

[20] Y. Tribis, A. El Bouchti and H. Bouayad, "Supply Chain Management based on Blockchain: A Systematic Mapping Study," *MATEC Web of Conferences,* vol. 200, 2018.

[21] Y. Chang, E. Iakovou and W. Shi, "Blockchain in Global Supply Chains and Cross Border Trade: A Critical Synthesis of the State-of-the-Art, Challenges and Opportunities," in *arXiv:1901.02715*, 2019.

[22] A. Kamilaris, A. Fonts and F. X. Prenafeta-Boldú, "The Rise of Blockchain Technology in Agriculture and Food Supply Chains," *Trends in Food Science & Technology Journal,* vol. 91, no. 1, pp. 640-652, 2019.

[23] F. Antonucci, S. Figorilli, C. Costa, F. Pallottino, L. Raso and P. Menesatti, "A Review on blockchain applications in the agri-food sector," *Journal of the Science of Food and Agriculture,* 2019.

[24] O. Bermeo-Almeida, M. Cardenas-Rodriguez, T. Samaniego-Cobo, E. Ferruzola-Gómez, R. Cabezas-Cabezas and W. Bazán-Vera, "Blockchain in agriculture: a systematic literature review," in *International Conference on Technologies and Innovation*, 2018.

[25] ICT4Ag, "Perspectives for ICT and Agribusiness in ACP countries: Start-up financing, 3D printing and blockchain," 2017. [Online]. Available: http://www.fao.org/e-agriculture/events/cta-workshop-perspectives-ict-and-agribusiness-acp-countries-start-financing-3d-printing-and.

[26] AgriDigital, 2017. [Online]. Available: https://www.agridigital.io/blockchain.

[27] A. Hoffman and R. Munsterman, "Dreyfus Teams With Banks for First Agriculture Blockchain Trade," 2018. [Online]. Available: https://www.bloomberg.com/news/articles/2018-01-22/dreyfus-teams-with-banks-for-first-agriculture-blockchain-trade.

[28] AID Tech, "How blockchain technology is enabling international aid to be delivered transparently," 2017. [Online]. Available: https://v3.aid.technology/aid-in-lebanon/.

[29] Blockchain for Zero Hunger, 2017. [Online]. Available: https://innovation.wfp.org/project/building-blocks.

[30] Ethereum, "The Ethereum Project," 2015. [Online]. Available: https://www.ethereum.org.

[31] Built to Adapt, "Can Blockchain Help Feed the Hungry?," 2018. [Online]. Available: https://builttoadapt.io/can-blockchain-help-feed-the-hungry-161b8fef7e24.

[32] M. Creydt and M. Fischer, "Blockchain and more-Algorithm driven Food Traceability," *Food Control,* 2019.

[33] CDC, "Centers for Disease Control and Prevention," 2018. [Online]. Available: https://www.cdc.gov/foodborneburden/2011-foodborne-estimates.html.

[34] Oceana, "Oceana Study Reveals Seafood Fraud Nationwide," 2013. [Online]. Available: http://oceana.org/reports/oceana-study-reveals-seafood-fraud-nationwide.

[35] H. Lee, H. Mendelson, S. Rammohan and A. Srivastava, "Technology in Agribusiness: Opportunities to drive value," White paper, Stanford Graduate School of Business, 2017.

[36] CB Insights, "How Blockchain Could Transform Food Safety," 2017. [Online]. Available: https://www.cbinsights.com/research/blockchain-grocery-supply-chain/.

[37] R. Kamath, "Food traceability on blockchain: Walmart's pork and mango pilots with IBM," *The JBBA,* vol. 1, no. 1, p. 3712, 2018.

[38] S. Wass, "Global Trade Review," 2017. [Online]. Available: https://www.gtreview.com/news/fintech/food-companies-unite-to-advance-blockchain-for-supply-chain-traceability/.

[39] CyberSecurity, "Milk Verification Project," 2019. [Online]. Available: https://www.greatitalianfoodtrade.it/consum-attori/blockchain-nella-filiera-alimentare-il-prototipo-di-bari.

[40] J. Kasten, "Blockchain Application: The Dairy Supply Chain," *Journal of Supply Chain Management Systems,* vol. 8, no. 1, 2019.

[41] F. Tian, "A supply chain traceability system for food safety based on HACCP, blockchain & Internet of things," in *International Conference on Service Systems and Service Management (ICSSSM)*, 2017.

[42] Zeto, "ZetoChain," 2018. [Online]. Available: https://www.zeto.ie/.

[43] T. Mohan, "Improve Food Supply Chain Traceability using Blockchain," Doctoral Dissertation, ed. by The Pennsylvania State University. State College, PA , Pennsylvania, USA, 2018.

[44] R. George, H. Harsh, P. Ray and A. Babu, "Food quality traceability prototype for restaurants using blockchain and food quality data index," *Journal of Cleaner Production,* vol. 240, no. 1, p. 118021, 2019.

[45] D. Tse, B. Zhang, Y. Yang, C. Cheng and H. Mu, "Blockchain application in food supply information security," in *International Conference on Industrial Engineering and Engineering Management (IEEM)*, 2017.

[46] J. Galvez, J. Mejuto and J. Simal-Gandara, "Future challenges on the use of blockchain for food traceability analysis," *TrAC Trends in Analytical Chemistry,* 2018.

[47] T. Levitt, "Blockchain technology trialled to tackle slavery in the fishing industry," 2016. [Online]. Available: https://www.theguardian.com/sustainable-business/2016/sep/07/blockchain-fish-slavery-free-seafood-sustainable-technology.

[48] MarketsandMarkets Research , "Food Traceability Market worth $14 Billion by 2019," 2016. [Online]. Available: https://www.marketsandmarkets.com/PressReleases/food-traceability.asp.

[49] J. Bunge, "Latest Use for a Bitcoin Technology: Tracing Turkeys From Farm to Table," The Wall Street Journal, 2017.

[50] Hendrix Genetics, "New blockchain project involving turkeys and animal welfare," 2018. [Online]. Available: https://www.hendrix-genetics.com/en/news/new-blockchain-project-involving-turkeys-and-animal-welfare/.

[51] Carrefour, "The Food Blockchain," https://actforfood.carrefour.com/Why-take-action/the-food-blockchain, 2018.

[52] Ireland Craft Beers, "Downstream beer," 2017. [Online]. Available: http://www.down-stream.io.

[53] Foodchain, "San Domenico roastery: The first case of a coffee supply chain fully traced with blockchain," 2019. [Online]. Available: https://food-chain.it/public/case/san-domenico/.

[54] Aldo Cozzi, "BlockChain arrives in the food sector: the packaging of pasta carries all its history with it," 2019. [Online]. Available: https://www.aldocozzi.it/news/la-blockchain-sbarca-nel-settore-alimentare-la-confezione-di-pasta-porta-con-se-tutta-la-sua-storia/ .

[55] A. Campbell, "Sustainability from paddock to plate," 2017. [Online]. Available: https://www.sciencealert.com/sustainability-from-paddock-to-plate.

[56] BeefLedger Limited, 2017. [Online]. Available: http://beefledger.io.

[57] JD.com Blog, "JD Blockchain Open Platform," 2018. [Online]. Available: https://jdcorporateblog.com/jd-launches-blockchain-open-platform/.

[58] Adele Peter, Fast Company, "In China, You Can Track Your Chicken On–You Guessed It– The Blockchain," 2017. [Online]. Available: https://www.fastcompany.com/40515999/in-china-you-can-track-your- chicken-on-you-guessed-it-the-blockchain.

[59] Grass Roots Farmers' Cooperative, "How we 're using blockchain tech for total transparency," 2017. [Online]. Available: https://grassrootscoop.com/blog/how-we-use-blockchain-technology-to-give-you-total-transparency.

[60] Hyperledger, "Bringing traceability and accountability to the supply chain through the power of Hyperledger Sawtooth's distributed ledger technology," 2018. [Online]. Available: https://sawtooth.hyperledger.org/examples/seafood.html.

[61] M. Caro, M. Ali, M. Vecchio and R. Giaffreda, "Blockchain-based traceability in Agri-Food supply chain management: A practical implementation," in *IoT Vertical and Topical Summit on Agriculture-Tuscany (IOT Tuscany)*, Tuscany, Italy, 2018.

[62] P. Suprunov, "Medium.com," 2018. [Online]. Available: https://medium.com/practical-blockchain/5-hyperledger-projects-in-depth-3d14c41f902b.

[63] D. Bechtsis, N. Tsolakis, A. Bizakis and D. Vlachos, "A Blockchain Framework for Containerized Food Supply Chains," *Computer Aided Chemical Engineering,* vol. 46, no. 1, pp. 1369-1374, 2019.

[64] B. Basnayake and C. Rajapakse, "A Blockchain-based decentralized system to ensure the transparency of organic food supply chain," in *International Research Conference on Smart Computing and Systems Engineering (SCSE)*, Colombo, Sri Lanka, 2019.

[65] K. Salah, N. Nizamuddin, R. Jayaraman and M. Omar, "Blockchain-based Soybean Traceability in Agricultural Supply Chain," *IEEE Access,* vol. 7, no. 1, pp. 73295 - 73305, 2019.

[66] WWF, "New Blockchain Project has Potential to Revolutionise Seafood Industry," 2018. [Online]. Available: https://www.wwf.org.nz/media_centre/?uNewsID=15541.

[67] Balfegó Group, 2017. [Online]. Available: https://balfego.com/ca/trasabilitat/.

[68] Ripe.io, 2017. [Online]. Available: http://www.ripe.io/.

[69] AgFunder News, "Maersk Leads Blockchain of Food Startup Ripe.io $2.4m Seed Round," 2018. [Online]. Available: https://agfundernews.com/maersk-leads-blockchain-of-food-startup-ripeio-2-4m-seed-round.html.

[70] OriginTrail, 2018. [Online]. Available: https://origintrail.io/use-cases.

[71] L. Ge, C. Brewster, J. Spek, A. Smeenk, J. Top, F. van Diepen, B. Klaase, C. Graumans and M. de Wildt, "Blockchain for agriculture and food," Wageningen Economic Research, No. 2017-112, Wageningen, 2017.

[72] A. S. Patil, B. A. Tama, Y. Park and K. H. Rhee, "A Framework for Blockchain Based Secure Smart Green House Farming," in *Advances in Computer Science and Ubiquitous Computing*, Singapore, 2017.

[73] ITUNews, "'Food Trust' partnership uses blockchain to increase food safety," 2018. [Online]. Available: https://news.itu.int/food-trust-blockchain-food-safety/.

[74] F. Tian, "An agri-food supply chain traceability system for China based on RFID & blockchain technology," in *13th International Conference on Service Systems and Service Management (ICSSSM)*, 2016.

[75] M. Kim, B. Hilton, Z. Burks and J. Reyes, "Integrating Blockchain, Smart Contract-Tokens, and IoT to Design a Food Traceability Solution," in *9th Annual Information Technology, Electronics and Mobile Communication Conference (IEMCON)*, 2018.

[76] V. Boehm, J. Kim and J. Hong, "Holistic tracking of products on the blockchain using NFC and verified users," in *International Workshop on Information Security Applications*, 2017.

[77] B. Abelseth, "Blockchain Tracking and Cannabis Regulation: Developing a permissioned blockchain network to track Canada's cannabis supply chain," *Dalhousie Journal of Interdisciplinary Management,* vol. 14, 2018.

[78] S. Figorilli, F. Antonucci, C. Costa, F. Pallottino, L. Raso, M. Castiglione, E. Pinci, D. Del Vecchio, G. Colle, A. Proto and G. Sperandio, "A blockchain implementation prototype for the electronic open source traceability of wood along the whole supply chain," *Sensors,* vol. 18, no. 9, p. 3133, 2018.

[79] M. Chinaka, "Blockchain technology - Applications in improving financial inclusion in developing economies: case study for small scale agriculture in Africa," Doctoral dissertation, Massachusetts Institute of Technology, 2016.

[80] FarmShare, 2017. [Online]. Available: http://farmshare.org.

[81] AgriLedger, 2017. [Online]. Available: http://www.agriledger.com/.

[82] D. Davcev, L. Kocarev, A. Carbone, V. Stankovski and K. Mitresk, "Blockchain-based Distributed Cloud/Fog Platform for IoT Supply Chain Management," in *Eighth International Conference On Advances in Computing, Electronics and Electrical Technology (CEET)*, 2018.

[83] OlivaCoin, 2016. [Online]. Available: http://olivacoin.com/.

[84] J. Bolt, "Financial resilience of Kenyan smallholders affected by climate change, and the potential for blockchain technology," Wageningen Environmental Research, Netherlands, 2019.

[85] Soil Association Certification, 2018. [Online]. Available: https://www.soilassociation.org/certification/.

[86] P. Lucena, A. Binotto, F. Momo and H. Kim, "A case study for grain quality assurance tracking based on a Blockchain business network," *arXiv preprint arXiv:1803.07877,* 2018.

[87] M. Chong, E. Perez, J. Castilla and H. Rosario, "Blockchain Technology Applied to the Cocoa Export Supply Chain: A Latin America Case," in *Handbook of Research on Emerging Technologies for Effective Project Management*, IGI Global, 2020, pp. 323-339.

[88] S. Jha, B. Andre and J. O., "ARBOL: Smart Contract Weather Risk Protection for Agriculture," 2018. [Online]. Available: https://www.arbolmarket.com/wp-content/uploads/2019/02/ARBOL_WP-3.pdf.

[89] ArbolMarket, "ARBOL," 2019. [Online]. Available: https://www.arbolmarket.com.

[90] H. Kim and M. Laskowski, "Agriculture on the blockchain: Sustainable solutions for food, farmers, and financing," in *Supply Chain Revolution*, Barrow Books, 2018.

[91] Plastic Bank, 2019. [Online]. Available: https://www.plasticbank.com.

[92] K. Steenmans and P. Taylor, "A rubbish idea: how blockchains could tackle the world's waste problem," 2018. [Online]. Available: https://theconversation.com/a-rubbish-idea-how-blockchains-could-tackle-the-worlds-waste-problem-94457.

[93] Agora Tech Lab, "Creating circular economies by rewarding responsible behavior," 2018. [Online]. Available: https://www.agoratechlab.com/about.

[94] SNCF, "How Blockchain simplifies waste disposal," 2017. [Online]. Available: https://www.digital.sncf.com/actualites/data-tritus-comment-la-blockchain-simplifie-le-tri-des-dechets.

[95] Recereum, 2017. [Online]. Available: http://recereum.com.

[96] Swachhcoin, "Decentralized Waste Management System," 2018. [Online]. Available: https://swachhcoin.com.

[97] D. Zhang, "Application of Blockchain Technology in Incentivizing Efficient Use of Rural Wastes: A case study on Yitong System," *Energy Procedia,* vol. 158, no. 1, pp. 6707-6714, 2019.

[98] H. Tao and D. Bullock, " Using Digital Agriculture Technologies to Improve Nitrogen Management and Wheat Yield," *Cereal Foods World,* vol. 64, no. 6, 2019.

[99] J. Lin, Z. Shen, A. Zhang and Y. Chai, "Blockchain and IoT based food traceability for smart agriculture," in *Proc. of the 3rd International Conference on Crowd Science and Engineering (ICCSE)*, Singapore, 2018.

[100] S. Keesstra, J. Bouma, J. Wallinga, P. Tittonell, P. Smith, A. Cerdà, L. Montanarella, J. Quinton, Y. Pachepsky, W. Van Der Putten and R. Bardgett, "The significance of soils and soil science towards realization of the United Nations Sustainable Development Goals," *Soil,* 2016.

[101] S. Keesstra, G. Mol, J. de Leeuw, J. Okx, M. de Cleen and S. Visser, "Soil-related sustainable development goals: Four concepts to make land degradation neutrality and restoration work," *Land,* vol. 7, no. 4, p. 133, 2018.

[102] R. Casado-Vara, J. Prieto, F. De la Prieta and J. Corchado, "How blockchain improves the supply chain: Case study alimentary supply chain," *Procedia computer science,* vol. 134, no. 1, pp. 393-398, 2018.

[103] M. Kouhizadeh, Q. Zhu and J. Sarkis, "Blockchain and the circular economy: potential tensions and critical reflections from practice," *Production Planning & Control,* pp. 1-7, 2019.

[104] D. Mao, F. Wang, Z. Hao and H. Li, "Credit evaluation system based on blockchain for multiple stakeholders in the food supply chain," *International journal of environmental research and public health,* vol. 15, no. 8, p. 1627, 2018.

[105] A. Pinna and S. Ibba, "A blockchain-based Decentralized System for proper handling of temporary Employment contracts," in *Science and Information Conference*, 2018.

[106] Gertrude Chavez-Dreyfuss, Reuters, "Coca-Cola, U.S. State Dept to use blockchain to combat forced labor," 2018. [Online]. Available: https://www.reuters.com/article/us-blockchain-coca-cola-labor/coca-cola-u-s-state-dept-to-use-blockchain-to-combat-forced-labor-idUSKCN1GS2PY?mc_cid=d7c996d219&mc_eid=4c123074ea.

[107] B. Bordel, D. Martín, R. Alcarria and T. Robles, "A Blockchain-based Water Control System for the Automatic Management of Irrigation Communities," in *Proc. of International Conference on Consumer Electronics (ICCE)*, Taiwan, 2019.

[108] M. Munir, I. Bajwa and S. Cheema, "An intelligent and secure smart watering system using fuzzy logic and blockchain," *Computers & Electrical Engineering,* vol. 77, no. 1, pp. 109-119, 2019.

[109] D. Brooker, F. Bakker-Arkema and C. Hall, Drying and storage of grains and oilseeds, Springer Science & Business Media., 1992.

[110] M. Kumar and N. Iyengar, "A framework for Blockchain technology in rice supply chain management," *Advanced Science Technology Letters,* vol. 146, pp. 125-130, 2017.

[111] F. Branco, F. Moreira, J. Martins, M. Au-Yong-Oliveira and R. Gonçalves, "Conceptual Approach for an Extension to a Mushroom Farm Distributed Process Control System: IoT and Blockchain," in *Proc. of the World Conference on Information Systems and Technologies (WorldCist)*, La Toja Island, Galicia, Spain, 2019.

[112] A. Poberezhna, "Addressing Water Sustainability With Blockchain Technology and Green Finance," in *Transforming Climate Finance and Green Investment with Blockchains*, 2018.

[113] A. Shaji, A. Rony, A. Kuriakose and F. Rawther, "Decentralized Land Lending System using Blockchain," *International Journal of Information,* vol. 8, no. 2, 2019.

[114] K. Behnke and M. Janssen, "oundary conditions for traceability in food supply chains using blockchain technology," *International Journal of Information Management,* 2019.

[115] H. Yuan, H. Qiu, Y. Bi, S. Chang and A. Lam, "Analysis of coordination mechanism of supply chain management information system from the perspective of block chain," *Information Systems and e-Business Management,* pp. 1-23, 2019.

[116] S. Pearson, D. May, G. Leontidis, M. Swainson, S. Brewer, L. Bidaut, J. Frey, G. Parr, R. Maull and A. Zisman, "Are Distributed Ledger Technologies the panacea for food traceability?," *Global Food Security,* vol. 20, pp. 145-149, 2019.

[117] Y. P. Lin, J. R. Petway, J. Anthony, H. Mukhtar, S. W. Liao, C. F. Chou and Y. F. Ho, "Blockchain: The Evolutionary Next Step for ICT E-Agriculture," *Environments,* vol. 4, no. 3, p. 50, 2017.

[118] Z. W. W. Li, G. Liu, L. Liu, J. He and G. Huang, "Toward open manufacturing: A cross-enterprises knowledge and services exchange framework based on blockchain and edge computing," *Industrial Management & Data Systems,* vol. 118, no. 1, pp. 303-320, 2018.

[119] E. C. Ferrer, "The Blockchain: A New Framework for Robotic Swarm Systems," in *Proceedings of the Future Technologies Conference*, 2018.

[120] M. Heinrich, F. Scotti, A. Booker, M. Fitzgerald, K. Kum and K. Löbel, "Unblocking high-value botanical value chains: Is there a role for blockchain systems?," *Frontiers in pharmacology,* vol. 10, no. 1, p. 396, 2019.

[121] A. Rejeb, "Blockchain Potential in Tilapia Supply Chain in Ghana," *Acta Technica Jaurinensis,* vol. 11, no. 2, pp. 104-118, 2018.

[122] J. Thomason, M. Ahmad, P. Bronder, E. Hoyt, S. Pocock, J. Bouteloupe, K. Donaghy, D. Huysman, T. Willenberg, B. Joakim and L. Joseph, "Blockchain—

Powering and Empowering the Poor in Developing Countries," *Transforming Climate Finance and Green Investment with Blockchains,* pp. 137-152, 2018.

[123] IWA, "Demystifying Blockchain for Water Professionals," 2018. [Online]. Available: http://www.iwa-network.org/demystifying-blockchain-for-water-professionals-part-1/.

[124] K. N. Khaqqi, J. J. Sikorski, K. Hadinoto and M. Kraft, "ncorporating seller/buyer reputation-based system in blockchain-enabled emission trading application," *Applied Energy,* vol. 209, pp. 8-19, 2018.

[125] F. Sander, J. Semeijn and D. Mahr, "The acceptance of blockchain technology in meat traceability and transparency," *British Food Journal,* vol. 120, no. 9, pp. 2066-2079, 2018.

[126] G. Perboli, S. Musso and M. Rosano, "Blockchain in Logistics and Supply Chain: A Lean Approach for Designing Real-World Use Cases," *IEEE Access,* vol. 6, pp. 62018-62028, 2018.

[127] K. Hald and A. Kinra, "How the blockchain enables and constrains supply chain performance," *International Journal of Physical Distribution & Logistics Management,* 2019.

[128] G. Zhao, S. Liu, C. Lopez, H. Lu, S. Elgueta, H. Chen and B. Boshkoska, "Blockchain technology in agri-food value chain management: A synthesis of applications, challenges and future research directions," *Computers in Industry,* vol. 109, pp. 83-99, 2019.

[129] S. Linsner, F. Kuntke, G. Schmidbauer-Wolf and C. Reuter, "Blockchain in Agriculture 4.0 - An Empirical Study on Farmers Expectations towards Distributed Services based on Distributed Ledger Technology," in *Proc. of Mensch und Computer (MUC)*, Hamburg, Germany, 2019.

[130] K. Rabah, "Convergence of AI, IoT, big data and blockchain: a review," *The Lake Institute Journal,,* vol. 1, no. 1, pp. 1-18, 2018.

[131] I. Mistry, S. Tanwar, S. Tyagi and N. Kumar, "Blockchain for 5G-enabled IoT for industrial automation: A systematic review, solutions, and challenges," *Mechanical Systems and Signal Processing,* vol. 135, no. 1, p. 106382, 2020.

[132] M. Swan and P. De Filippi, "Toward a Philosophy of Blockchain: A Symposium: Introduction," *Metaphilosophy,* vol. 48, no. 5, pp. 603-619, 2017.

[133] 1000EcoFarms, 2017. [Online]. Available: https://www.1000ecofarms.com/.

[134] FoodCoin, "The FoodCoin Ecosystem," 2017. [Online]. Available: https://www.foodcoin.io/.

[135] X. Xu, Q. Lu, Y. Liu, L. Zhu, H. Yao and A. V. Vasilakos, "Designing blockchain-based applications a case study for imported product traceability," *Future Generation Computer Systems,* vol. 92, pp. 399-406, 2019.

[136] J. Yli-Huumo, D. Ko, S. Choi, S. Park and K. Smolander, "Where is current research on blockchain technology?—a systematic review," *PloS one,* vol. 11, no. 10, 2016.

[137] S. Gaurav, "The Market for Cryptocurrencies," *Economic & Political Weekly,* vol. 54, no. 2, p. 13, 2019.

[138] P. Jayachandran, "The difference between public and private blockchain," 2017. [Online]. Available: https://www.ibm.com/blogs/blockchain/2017/05/the-difference-between-public-and-private-blockchain/.

[139] S. Chen, R. Shi, Z. Ren, J. Yan, Y. Shi and J. Zhang, "A blockchain-based supply chain quality management framework," in *IEEE 14th International Conference on e-Business Engineering (ICEBE)*, Shanghai, China, 2017.

[140] I. Eyal, A. Gencer, E. Sirer and R. Van Renesse, "Bitcoin-ng: A scalable blockchain protocol," in *Proceedings of the 13th USENIX Symposium on Networked Systems Design and Implementation (NSDI)*, Santa Clara, CA, USA, 2016.

[141] D. Mao, Z. Hao, F. Wang and H. Li, "Innovative Blockchain-Based Approach for Sustainable and Credible Environment in Food Trade: A Case Study in Shandong Province, China," *Sustainability,* vol. 10, no. 9, p. 3149, 2018.

[142] Q. Feng, D. He, S. Zeadally, M. Khan and N. Kumar, "A survey on privacy protection in blockchain system," *Journal of Network and Computer Applications,* 2018.

[143] H. Kim and M. Laskowski, "Toward an ontology-driven blockchain design for supply-chain provenance," *Intelligent Systems in Accounting, Finance and Management,* vol. 25, no. 1, pp. 18-27, 2018.

[144] W. E., "Putting Things to REST," UCB iSchool Report 2007-015, Berkeley, USA, 2007.

[145] A. Kamilaris, A. Pitsillides and V. Trifa, "The smart home meets the web of things," *International Journal of Ad Hoc and Ubiquitous Computing,* vol. 7, no. 3, pp. 145-154, 2011.

[146] A. Maru, D. Berne, J. Beer, P. Ballantyne, V. Pesce, S. Kalyesubula, N. Fourie, C. Addison, A. Collett and J. Chavez, "Digital and data-driven agriculture: Harnessing the power of data for smallholders," *Global Forum on Agricultural Research and Innovation,* 2018.

[147] A. Kamilaris, A. Kartakoullis and F. X. Prenafeta-Boldú, "A review on the practice of big data analysis in agriculture," *Computers and Electronics in Agriculture,* vol. 143, pp. 23-37, 2017.

[148] J. Killmeyer, M. White and B. Chew, "Will blockchain transform the public sector?," Deloitte Center for Government Insights, Deloitte University Press, 2017.

[149] V. Gupta, "Building the Hyperconnected Future on Blockchains," World Government Summit. http://internetofagreements.com/files/WorldGovernmentSummit- Dubai2017.pdf, 2017.